\documentclass[prd, twocolumn, nofootinbib, floatfix]{revtex4-1}

\usepackage{amsmath}
\usepackage{graphicx}
\usepackage{dcolumn}
\usepackage{bm}
\usepackage{epsfig}
\usepackage{amssymb,latexsym,mathrsfs}
\usepackage{graphicx}
\usepackage{color}
\usepackage{hyperref}

\newcommand{\be}{\begin{equation}}
\newcommand{\ee}{\end{equation}}
\newcommand{\bs}{\begin{split}} 
\newcommand{\bea}{\begin{eqnarray}}
\newcommand{\eea}{\end{eqnarray}}

\newcommand{\om}{\Omega_m}

\begin{document}

\title{Lensing Time Delays and Cosmological Complementarity} 
\author{Eric V.\ Linder} 
\affiliation{Berkeley Lab \& University of California, Berkeley, 
CA 94720, USA}
\affiliation{Institute for the Early Universe WCU, Ewha Womans 
University, Seoul, Korea}

\begin{abstract}
Time delays in strong gravitational lensing systems possess significant 
complementarity with distance measurements to determine the dark energy 
equation of state, as well as the matter density and Hubble constant.  Time 
delays are most useful when observations permit detailed lens modeling 
and variability studies, requiring high resolution imaging, long time 
monitoring, and rapid cadence.  We quantify the constraints possible between 
a sample of 150 such time delay lenses and a near term supernova program, 
such as might become available from an Antarctic telescope such as KDUST 
and the Dark Energy Survey.  Adding time delay data to supernovae plus 
cosmic microwave background information can improve the dark energy figure 
of merit by almost a factor 5 and determine the matter density $\Omega_m$ 
to 0.004, Hubble constant $h$ to 0.7\%, and dark energy equation of state 
time variation $w_a$ to 0.26, systematics permitting. 
\end{abstract}

\date{\today} 

\maketitle

\section{Introduction} 

Complementarity between cosmological probes increases their leverage on 
the cosmological model parameters, crosschecks results through differing 
systematic uncertainties, and breaks degeneracies.  These all play 
important roles in elucidating the nature of our universe: the energy 
densities in matter and dark energy, the scale of the universe through 
the Hubble constant, and the characteristics of the dark energy behind 
the current cosmic acceleration. 

Most probes, however, have substantial similarity in their parameter 
dependencies, involving the same combinations of ingredients entering 
into the Hubble parameter as a function of redshift, $H(z)$.  Distances 
(and volumes) in particular, are essentially equivalent, and growth of 
structure also depends similarly on $H(z)$.  Looking for a high degree 
of complementarity, especially to determine the dark energy equation of 
state value and time variation, \cite{linsl} investigated the use of 
distance ratios present in strong gravitational lensing as a means of 
breaking this degeneracy. 

While lensing distance ratios involve the mass structure of the 
lens, and so are not purely geometric, there has been impressive progress 
in modeling the mass distributions in lensing systems (e.g.\ 
\cite{lensmass1,lensmass2,fadely10}) 
and so it is worth considering strong lensing distances in more detail 
as a cosmological probe, in particular for its complementarity.  Here 
we revisit \cite{linsl} with several important distinctions: 1) we 
concentrate on time delays, due to the recent observational successes 
\cite{suyu10,fadely10} and modeling advances; 2) we consider a more 
realistic range of future observational prospects, involving projects 
starting to get underway, which will have important implications for 
complementarity; and 3) we carry out studies of the science 
reach as a function of redshift range, and in the presence of spatial 
curvature. 

Several authors 
(e.g.\ \cite{oguri,coe,koopmans09,dobke,oguri07,lewisibata,yamamoto}) 
have addressed the statistical power 
of strong lensing time delays from further future surveys such as LSST, 
calculating the numbers of lenses found and with measured time delays, 
and projecting possible Hubble constant or cosmological constraints.  
These, however, treat 
the lensing systems as an ensemble to average over, and in fact identify 
the mass modeling as a major uncertainty capable of degrading constraints 
substantially.  Here we concentrate on what are sometimes called ``golden 
lenses'', although now the meaning is not systems with some special 
symmetry but rather ones where the survey design has specifically provided 
data enabling detailed construction of the lens mass model.  The number 
of such systems will be much less but we find they can have significant 
scientific leverage. 

In Section~\ref{sec:method} we discuss the cosmological impact 
of time delay measurements and their complementarity with other probes. 
Section~\ref{sec:results} considers reasonable possibilities for survey 
data sets in terms of number and redshift range of time delay systems, 
and analyzes their constraints in conjunction with a mid term supernova 
survey and cosmic microwave background data.  Survey requirement issues 
with respect to imaging resolution, time sampling, etc.\ for time delay 
measurement, lens modeling, and systematic error control are outlined in 
Section~\ref{sec:surveys}, with specific reference to the Antarctic 
optical/infrared telescope program.

\section{Time Delays as Cosmic Probe} \label{sec:method} 

Strong gravitational lensing causes multiple images of distant sources, 
with the light rays from the images taking different amounts of time to 
propagate to the observer.  The time delay involves two parts: a geometric 
delay from different path lengths and a gravitational time delay from 
traversing different values of the gravitational potential of the lens. 
Thus both the image positions and the lens mass model must be accounted for. 
Time delays are observed by looking for coordinated variations in the 
flux from the images, e.g.\ of time varying quasars, which requires long 
time, well sampled monitoring.  
Typical galaxy lens induced delays are $\sim60$ days and the desired 
measurement accuracy is a couple of days or better. 

To translate the image angular positions into spatial positions, for 
computing both the path length and the gravitational potential effects, 
one needs the (conformal) distance to the lens, $r_l$, to the source, 
$r_s$, and between the source and lens, $r_{ls}$.  Only in flat space is 
$r_{ls}=r_s-r_l$.  The particular combination of distances central to 
time delays is 
\be 
T\equiv \frac{r_l r_s}{r_{ls}} \,. 
\ee 

Specifically, following \cite{suyu10}, the time delay of an image at 
position $\vec\theta$ on the sky relative to an unlensed source at position 
$\vec\beta$ is 
\be 
\Delta t(\vec\theta,\vec\beta)=\frac{r_l r_s}{r_{ls}}\,(1+z_l)\, 
\phi(\vec\theta,\vec\beta) \,, 
\ee  
where the distance ratio is $T$, containing the key cosmological 
information, and $\phi$ is the Fermat or time delay potential given by 
\be  
\phi(\vec\theta,\vec\beta)=\frac{(\vec\theta-\vec\beta)^2}{2}-\psi(\vec\theta) 
\,. 
\ee 
We see that the first term on the right hand side 
is the geometric delay and the second 
term is the lensing potential delay with $\nabla^2\psi=2\kappa$ for 
$\kappa$ the dimensionless lensing projected surface mass density. 

The time delay potential $\phi$ connecting $T$ to the observed time delays 
therefore depends on the 
lens mass distribution, the model for which is built up from information 
on image positions and flux ratios and perhaps surface brightness 
morphologies, ideally from many images including 
arcs.  See \cite{suyu10,fadely10,oguri07} for details on the modeling process 
and calculation of the  potential factor.  All the cosmological information 
comes from $T$ (cf.\ the approach of \cite{hjorth}), with the 
uncertainties in the  potential factor entering into the 
error propagation, together with measurement uncertainties. 

The time delay probe $T$ has interesting properties with regards to the 
cosmological parameter leverages.  As noted by \cite{linsl}, the 
sensitivities to dark energy parameters $w_0$ and $w_a$, where the 
dark energy equation of state is well fit by $w(z)=w_0+w_a z/(1+z)$ 
\cite{linprl}, are actually positively correlated in contrast to standard 
distance measurements such as from Type Ia supernovae.  This offers hope 
for complementarity with such probes.  Furthermore, the sensitivity to 
the dimensionless matter density $\om$ at low redshift is remarkably low 
compared to solo distances, leading to the possibility of breaking the 
usual degeneracy between matter density and dark energy equation of state.  
Finally, the ratio depends linearly on the Hubble scale $H_0^{-1}$, and 
since \cite{refsdal} researchers have sought to use lensing time delays to 
measure the Hubble constant. 

\begin{figure}[htbp!]
\includegraphics[width=\columnwidth]{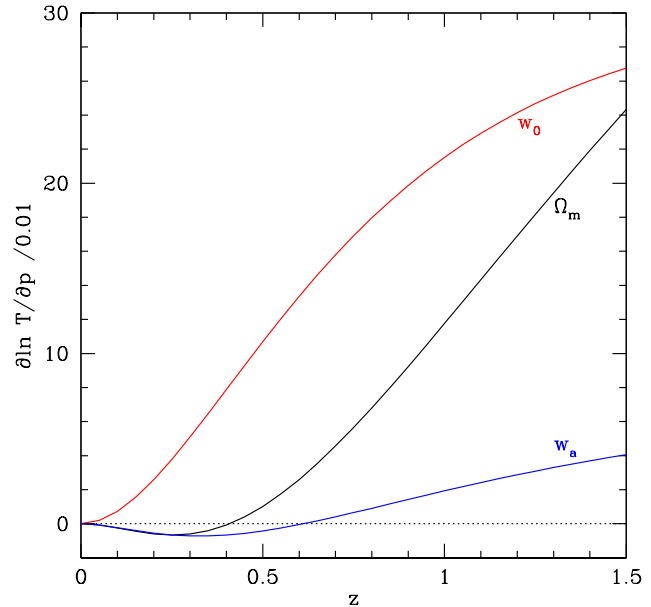}
\caption{Sensitivity of the time delay distance combination 
$T=r_l r_s/r_{ls}$ to the cosmological parameters is plotted vs lens 
redshift.  Curves with opposite signs at the same redshift indicate 
positive correlations between those parameters -- very unusual for 
the dark energy equation of state variables $w_0$ and $w_a$.  
}
\label{fig:sens}
\end{figure}

In Figure~\ref{fig:sens} we highlight 
these special properties.  The derivatives $\partial\ln T/\partial p$ 
give the sensitivities for each parameter $p$ and are exactly what enters 
into a Fisher matrix analysis for cosmological parameter estimation. 
Raw (unmarginalized) sensitivities can be read directly, e.g.\ 
$\partial\ln T/\partial p/0.01=10$ means that a 1\% measurement of $T$ 
delivers an uncertainty $\sigma(p)=0.1$.  This must be folded in with 
the covariances between parameters: sensitivity curves having the same 
(reflected) shape indicate highly anticorrelated (correlated) parameters. 
For the time delay probe, the curve shapes are not very similar -- a 
good sign for breaking degeneracies -- and we see the unusual positive 
correlation between $w_0$ and $w_a$ over the whole range $z_l=0$--0.6 
(where for simplicity we have assumed $z_s=2z_l$).

To take advantage of the odd correlation properties of $T$ to give 
strong complementarity in probing cosmology, we include Type Ia supernova 
distances as another probe, having very different degeneracies.  The 
supernova distance-redshift relation has no sensitivity to the Hubble 
constant $h=H_0/(100\,$km/s/Mpc) however, this being convolved with the 
unknown supernova absolute luminosity.  To profit from the time delay 
dependence on $h$, therefore, we also use cosmic microwave background (CMB) 
information, which determines the physical matter density combination 
$\om h^2$ very well (but not particularly $\om$ by itself).  This further 
offsets the weak dependence of $T$ alone on the matter density, and so the 
weakness of each is turned into strength in complementarity.

\section{Cosmological Leverage} \label{sec:results} 

Another interesting property of the time delay probe is that its 
useful and unusual correlation properties occur at low redshift, 
for $z_l=0$--0.6.  Detailed observations of lensing systems will 
be easier there, where the lens galaxy and source images will not 
be as faint as at higher redshift.  We therefore take as our baseline 
a survey producing time delay measurements at $z_l=0.1$--0.6 (there is 
relatively little volume for lensing systems below $z_l=0.1$), and then 
study variations of this.  For simplicity we fix $z_s=2z_l$; although 
there will be a distribution of source redshifts this has little impact 
on the cosmology estimation (see, e.g., \S5.5 of \cite{coe}) and we have 
explicitly checked that using instead $z_s=4z_l$ affects the dark energy 
figure of merit (uncertainty area) result by less than 1\%.  In most of 
this section we assume a spatially flat universe, studying the effect of 
an additional parameter for curvature in Section~\ref{sec:curv}.

\subsection{Cosmological Parameter Constraints} \label{sec:flat} 

To the time delay measurements we add supernova distance (SN) and CMB 
information and carry out a Fisher matrix analysis to estimate the 
cosmological parameter constraints.  For the supernovae, we take a 
mid term sample reasonable for the next five years, consisting of 
150 SN at $z=0.03$--0.1 from the Nearby Supernova Factory \cite{snf}, 
100 SN per 0.1 bin in redshift from $z=0.1$--1 as from the Dark Energy 
Survey (DES: \cite{des}) with follow up spectroscopy, and 42 SN between 
$z=1$--1.7 as from Hubble Space Telescope observations such as the 
CLASH \cite{clash} and CANDELS \cite{candels} surveys.  This seems like 
a reasonable estimate for a mid term, well characterized supernova sample. 
Each supernova is given a 0.15 mag (7\% in distance) statistical uncertainty 
and each redshift bin of 0.1 has a systematic floor at 
$dm_{\rm sys}=0.02\,(1+z)$ added in quadrature to the statistical error. 
Thus the supernova sample is systematics limited out to $z=1$.  
For CMB data, we take Planck quality information consisting of determination 
of the geometric shift parameter $R$ to 0.2\% and the physical matter density 
$\om h^2$ to 0.9\%, roughly corresponding to constraints from the location 
and amplitude, respectively, of the temperature power spectrum acoustic 
peaks.  The parameter set is $\{\om,w_0,w_a,h,{\mathcal M}\}$, where 
$\mathcal{M}$ is the convolution of the supernova absolute luminosity 
and the Hubble constant. 

Current measurements can deliver the time delay probe $T$ to $\sim5\%$ 
for a lensing system, dominated by systematic uncertainties for 
individual systems.  With a survey designed to find many strong lensing 
images and characterize them accurately, it may be possible to consider 
1\% measurements of $T$ in each redshift bin of 0.1 from $z=0.1$--0.6. 
This can be thought of as either 25 strong lenses per bin (150 total), 
or fewer lenses with better accuracy than 5\% per system 
from a survey designed to gather data needed to control systematics, or 
a combination of the two.  We discuss the survey requirements in 
Section~\ref{sec:surveys}. 

\begin{figure}[htbp!]
\includegraphics[width=\columnwidth]{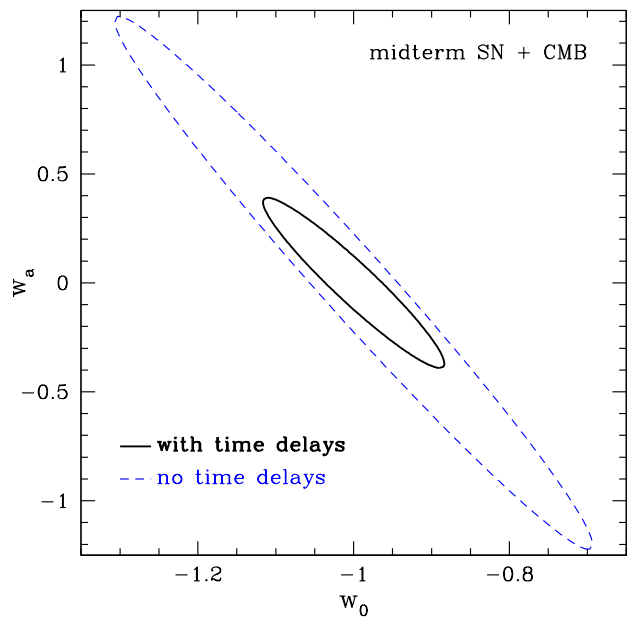}
\caption{68\% confidence level constraints on the dark energy equation 
of state parameters $w_0$ and $w_a$ using mid term supernova distances 
and CMB information, and with (solid curve) or without (dashed curve) 
time delay measurements.  The time delay probe demonstrates strong 
complementarity, tightening the area of uncertainty by a factor 4.8. 
}
\label{fig:t16}
\end{figure}

Figure~\ref{fig:t16} shows the dramatic improvement in the dark energy 
equation of state parameters (marginalized over the other parameters) 
when adding the time delay probes of 1\% accuracy over $z=0.1$--0.6.  
The area of the error contour in $w_0$--$w_a$ tightens by a factor 4.8 over 
that from SN+CMB alone.  All the cosmological parameters are better 
determined by factors of 2.6--3.1.  Time delays therefore have great 
complementarity with the supernova and CMB probes, and such a strong 
lensing survey would be highly valuable scientifically.

The absolute level of the constraints with time delays is impressive 
as well.  The Hubble constant is determined to 0.0051, or 0.7\%; the 
matter density $\om$ to 0.0044 (1.6\%), and the present value of the 
dark energy equation of state $w_0$ to 0.077 and its time variation 
$w_a$ to 0.26.  While falling short of the results from a space survey 
of supernovae (with CMB), such a mid term program could deliver important 
insights into the nature of cosmic acceleration and the cosmological model. 

The baseline time delay sample adopted seems plausible, but let us consider 
variations to see how the cosmological constraints depend on the survey 
characteristics.  It may be difficult to find enough 
strong lens systems at the lowest redshifts, due to the limited volume.  
Note however that the SLACS survey has been successful in detecting 
lenses \cite{slacs}, if not necessarily measuring time delays, at 
$z_l\approx0.1$, and this depends on the source population targeted. 
Nevertheless, if we cut the time delay information to the range 
$z=0.3$--0.6 (so 100 strong lenses), we find that this reduces the 
figure of merit (inverse uncertainty area) by 25\%.  
The greatest effect is on the Hubble constant determination, since this 
is what low redshift time delays excel at, with $\sigma(h)$ degrading 
by 55\%.  This then propagates into the $\om$ constraint, which weakens 
by 41\%.  These can be somewhat ameliorated if we have some information 
from $z=0.1$--0.2, e.g.\ using 12 rather than 50 time delays in this 
range recovers almost half the constraining power. 

Conversely, suppose that a strong lensing time delay survey could be 
extended out to $z_l=1$, still at the 1\% accuracy per 0.1 redshift bin. 
Then the figure of merit improves by 40\%, though the constraints on 
$\om$ and $h$ only gain by 6\%.  Detailed characterization of the lensing 
systems at such high redshift could be problematic, however, due to 
lower fluxes and signal to noise.  The redshifts of well characterized 
time delay systems is slowly being pushed out toward $z_l=1$ 
\cite{suyuprivate,cosmograil}.

\subsection{Including Curvature} \label{sec:curv}

Spatial curvature enters together with the Hubble parameter into either 
the angular or luminosity distance between observer and source.  
Degeneracy between the 
curvature density $\Omega_k=1-\Omega_m-\Omega_{de}$ and dark energy 
equation of state can be severe; for example see Fig.~6 of \cite{lincurv} 
for effects on $w_0$, $w_a$ or \cite{armandegen} for general $w(z)$.  
This can be broken by using a wide redshift range of distances; in 
particular high and low redshift distance measurements can separate the 
curvature density from other components.  Another possibility is direct 
measurement of the Hubble parameter as well as distances (e.g.\ from 
the radial baryon acoustic oscillation scale), or distance ratios appearing 
in gravitational lenses or large scale structure (see, e.g., 
\cite{knox,bern,linap}).  This has the 
advantage of not necessarily requiring high redshift measurements. 

\begin{figure}[htbp!]
\includegraphics[width=\columnwidth]{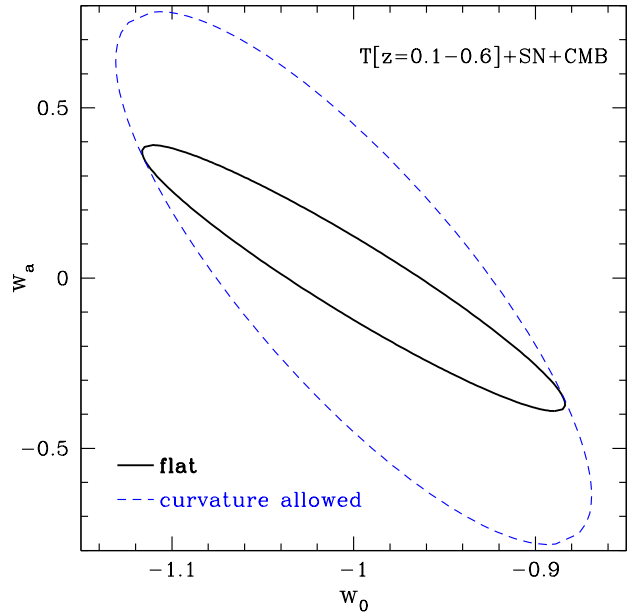}
\caption{68\% confidence level constraints on the dark energy equation
of state parameters $w_0$ and $w_a$ using time delay, mid term supernovae, 
and CMB information, assuming spatial flatness (solid curve) or allowing 
curvature (dashed curve).  The time delay probe moderates the curvature 
degeneracy, restricting the area degradation to a factor 4, rather than 20 
without time delay data. 
}
\label{fig:noflat}
\end{figure}

We now examine the role that time delay measurements can play in 
breaking the curvature degeneracy, if the universe is not assumed to 
be spatially flat.  Figure~\ref{fig:noflat} shows the results when we 
allow for curvature in the cosmology fitting, using time delays, 
supernovae distances, and CMB information.

The dark energy equation of state uncertainty indeed degrades, with the 
area figure of merit declining by a factor 4.1.  This shows the degeneracy 
is not fully broken, but should be contrasted with the factor 20.2 
degradation from including curvature with only the SN+CMB data for 
constraint.  Thus the time delay probe is a useful tool even/especially 
when allowing for spatial curvature.  Most of the covariance affects the 
time variation $w_a$, with its uncertainty doubling.  The present value 
$w_0$ is only determined 12\% worse, and the errors on $\om$ and $h$ 
increase by 29\% and 27\%.  The curvature itself is estimated to 
$\sigma(\Omega_k)=0.0063$.

\section{Survey Characteristics} \label{sec:surveys} 

In order to use time delays as a cosmological probe in the individual 
lensing system approach, the survey must deliver detections and accurate 
measurements of the time delays, detailed modeling of the lens systems, 
and control of other systematic uncertainties.  Systematics include 
microlensing that induces variability, differentially altering the images' 
light curves, and projected mass not truly part of the lens, altering the 
mass modeling. 

To detect a large sample of time delay systems, a wide field survey is 
needed, but to characterize them through accurate image positions, 
splittings, and flux variations requires high resolution imaging.  
Interestingly, 
a telescope at an excellent seeing site such as Dome A, Antarctica 
\cite{domea} could fulfill both roles.  The Kunlun Dark Universe 
Telescope (KDUST: \cite{kdust}), a 2.5 meter telescope planned for Dome A 
would be situated above the low ground layer and possibly have $0.3''$ 
median seeing in the optical, $0.2''$ in the low background noise infrared.

The advantages of high resolution imaging for strong lensing are crucial 
and manifold \cite{richard,koopmans09}.  Such seeing also helps to separate 
the images from contamination by the lens galaxy light.  To take advantage 
of this excellent seeing for strong lensing, the point spread function (PSF) 
would need to be stable, or algorithms developed to fit simultaneously the 
PSF and lens mass model.  The stable winter weather, with low winds and 
large isoplanatic angle, at Dome A could be advantageous.  KDUST surveys 
would overlap with Dark Energy Survey fields, as well as those of the 
South Pole Telescope and LSST.  DES could supply much of the supernova 
sample, although supernova programs, at either low or very high redshift, 
are also being studied for KDUST \cite{alexkdust}. 

Measuring time delays accurately from detected strong lensing systems 
requires a long time baseline, since the time delay distribution of 
interest is in the range of $\sim$10--100 days.  The long Antarctic night 
offers advantages here of continuity, although end effects from the long 
Antarctic day mean that not all systems at the upper end of this 
distribution will be usable, in particular the longer cluster lenses 
(which also likely have 
larger external mass effects).  Dense time sampling enables 
accurate time delay determination, and ameliorates the effect of 
microlensing systematics, and again the Antarctic site allows continuous 
viewing of fields and regular sampling, every 24 hours or even more often.  
(Indeed, detection of time delay perturbations can be used as a probe 
of dark matter substructure and properties \cite{keeton}.)  
While photometric redshifts 
are likely good enough to remove most projected mass contamination, 
follow up spectroscopy for accurate determination of the redshifts of the 
images and lens constituents is necessary.  A spectrographic telescope is 
being considered for Dome A, but spectroscopy is needed in any case for 
DES fields (e.g.\ for the supernovae), so KDUST gains further advantage 
from synergy with DES. 

The prospects for 1\% measurement of the time delay-redshift relation 
in a 0.1 bin in redshift for lenses at $z=0.1$--0.6 seem reasonable. 
Improvements in control of systematics could tighten the current 5\% 
accuracy, and such surveys will build the statistics as well.  The 
number of time delay systems baselined in this article -- 150, a one 
order of magnitude increase over current levels -- is plausible, as is the 
range of supernovae data, making the science case for time delay surveys 
of interest for further investigation.

\section{Conclusions} \label{sec:concl} 

We have quantified the significant complementarity as cosmological 
probes that strong gravitational lensing time delays, involving 
distance ratios, have with solo distance measurements such as from 
Type Ia supernovae.  A well designed time delay survey can add to 
practical, near term supernova and CMB data to provide surprisingly 
incisive constraints on the dark energy equation of state, the Hubble 
constant, and the matter density.  The improvement in equation of state 
area uncertainty (figure of merit) is almost a factor 5 over the data 
sets without time delays. 

Time delays also significantly ameliorate the degeneracies in parameter 
determination caused by allowing for spatial curvature, again improving 
the area uncertainty by a factor 5.  Determination of the Hubble constant 
to 0.7\% as well would be valuable for several astrophysical and 
cosmological applications.

We have focused on what seem to be near term, reasonable data sets. 
An exciting possibility for achieving these is telescopes being developed 
at promising Antarctic astronomical sites, such as KDUST at Dome A.  
If these truly deliver high resolution, stable seeing much better than 
conventional ground based optical conditions (if not quite space quality), 
the baseline time delay survey considered here to deliver one order of 
magnitude times larger sample of well characterized time delay systems 
appears practical.  Another advantage is the synergy with other southern 
surveys, such as the Dark Energy Survey in the near term.  (While we have 
intentionally not extrapolated to long term developments, synergy with 
LSST is clear as well.) 

Systematic uncertainties would be ameliorated by the high resolution 
imaging, whether single epoch to characterize in detail the lens model 
and separate the host galaxy light, or multiepoch to finely measure the 
flux variations and measure clean and accurate time delays.  The redshift 
range for the survey could be modest, $z_l\approx0.1$--0.6, and we 
presented how the cosmological constraints change if a narrower or wider 
range is considered. 

In attempting to stay within straightforward practicality we have 
not discussed exciting ideas such as testing for deviations from 
general relativity.  After all, the same principle is used in the 
solar system, with spacecraft signal time delays providing stringent 
limits on other gravity theories; also some screening mechanisms that 
restore gravity to general relativity are expected to kick in on scales 
accessible to cosmological strong lensing \cite{tristan,schmidt,jainkhoury}.  
Many other astrophysical 
applications exist for a high resolution imaging survey (especially with 
low noise in the infrared), such as using strong lenses as gravitational 
telescopes to study early structures and the epoch of reionization. 

Complementarity between cosmological probes offers the strongest and 
most robust leverage for revealing the scale and contents of our universe, 
and the nature of the cosmic acceleration.  The combination of time delays, 
supernova distances, and CMB data provides an exciting level of insight 
with near term surveys.

\acknowledgments

I gratefully acknowledge Sherry Suyu for helpful discussions, and thank 
the Niels Bohr International Academy and Dark Cosmology Centre, University 
of Copenhagen for hospitality during the workshop inspiring this research.  
This work has been supported in part by the Director, 
Office of Science, Office of High Energy Physics, of the U.S.\ Department 
of Energy under Contract No.\ DE-AC02-05CH11231 and by World Class 
University grant R32-2009-000-10130-0 through the National Research 
Foundation, Ministry of Education, Science and Technology of Korea.


\end{document}